\documentclass[aps,prl,twocolumn]{revtex4}
\usepackage{hyperref}
\usepackage{natbib}
\usepackage{graphicx}
\usepackage{amssymb, amsmath}
\usepackage[normalem]{ulem}

\def\rads{\hbox{\rm\hskip.35em rad s}$^{-1}$}

\def\gcs{\hbox{\rm\hskip.35em  g cm}$^{-2}$}
\hypersetup{colorlinks=true, citecolor=blue, linkcolor= magenta}
\usepackage[normalem]{ulem}

\begin{document}

\title{Nonlinear superfluidity and time-delay based chaotic spin-down in pulsars}
\author{Erbil G\"{u}gercino\u{g}lu}
\email{egugercinoglu@gmail.com}
\affiliation{Faculty of Engineering and Natural Sciences, Sabanc{\i} University, Orhanl{\i}, 34956 Istanbul, Turkey}
\author{Mustafa Do\u{g}an}
\email{mustafa.dogan@gmail.com}
\affiliation{Faculty of Electrical-Electronics Engineering, Department of Control and Automation Engineering, Istanbul Technical University, Maslak, 34469, Istanbul, Turkey}
\author{K. Yavuz Ek{\c s}i}
\email{eksi@itu.edu.tr}
\affiliation{Faculty  of Science  and  Letters, Physics Engineering  Department, Istanbul Technical University, Maslak, 34469, Istanbul, Turkey}


\begin{abstract}
We investigate the chaotic spin-down behavior seen from some pulsars in terms of the nonlinear superfluid dynamics. To this end, we numerically solve the set of equations for the superfluid-normal matter system whose coupling is mediated by creep of the vortex lines. We show that glitch perturbations which introduce a time-delay to the steady-state dynamics leave behind a remnant in the third time derivative of the rotational phase. This time-delay induces a hyper-chaotic spin-down for pulsars. We find that glitch-induced changes in the rotational parameters lead to non-closing cyclic patterns in the time-delayed phase difference diagram. We observe that the number of cycles, $N$, in the diagram results from $N+1$ glitches occurred in total observation time.
\end{abstract}

\pacs{97.60.Jd, 97.60.Gb, 05.45.−a, 74.25.Qt}

\maketitle

\textit{Introduction.--}
Pulsars, being rapidly rotating neutron stars, tend to be extremely stable rotators due to their compact sizes and huge anchored magnetic fields, and in some cases even rival the best terrestrial atomic clocks. Such a stable rotation enables one to predict the rotational evolution of a pulsar with high accuracy. Nevertheless, for some cases offsets in the form of noise as well as oscillatory behavior have been found in the long-term pulsar time-series data \cite{hobbs10,namkham19,parthasarathy19,lower20,dang20}. Lyne et al. \cite{lyne10} interpreted these anomalies in terms of magnetospheric switching between unspecified two different slow-down laws. In Ref. \cite{biryukov12} pulsar spin-down is considered as a combination of one monotonic component due to magneto-dipole radiation and other cyclic component whose underlying cause remained unanswered. Moreover, many young pulsars exhibit glitches, i.e.\ their rotation rate as well as spin-down rate suddenly increases \cite{espinoza11,yu13,erbil21}. These increments in the rotation and spin-down rates tend to recover gradually towards the original spin behavior. Sudden increases in spin parameters followed by slow relaxation imply that glitches originate from the interior superfluid component(s) loosely coupled to the observed neutron star crust \cite{baym69,alpar84}.      

An alternative explanation of the irregularities in the long-term pulsar spin-down behavior could be the display of chaotic behavior governed by the underlying dynamical system. 
Harding et al. \cite{harding90} investigated the chaotic behavior of the Vela pulsar through phase residuals and compared data with a combination of random walks in both frequency and frequency derivative. Seymour and Lorimer \cite{seymour13} searched for Lorenz and R\"{o}ssler type chaotic attractors in the spin-down rates of 17 pulsars in Ref. \cite{lyne10}. 

By examining time-delayed spin-down rates non-closing orbits were detected in PSR B1540--06, PSR B1828--11, PSR B1826--17 and PSR B1642--03  \cite{seymour13}. All of these features are indicators of chaotic imprints in pulsar spin-down. It was also noted that detailed long-term time-series analysis of other pulsars may reveal underlying chaotic behavior in future observations \cite{seymour13}.  

Classical chaotic attractors would require that at least three-dimensional coupled dynamical equation system is governing the rotational evolution of pulsars \cite{harding90,seymour13}. The algorithm used in Ref. \cite{seymour13} is sensitive to low-dimensional chaos such that Lyapunov exponents are bounded to small values due to limited data. We note, however, that chaotic behavior can be exhibited even in a first-order system if the rate of change of a quantity depends on the value of that quantity at an earlier time \cite{mackey77}. The existence of unknown long-enough time-delay in a system makes it infinite-dimensional and results in oscillatory behavior leading to so-called hyper-chaos \cite{pietri09,kuptsov20,junges+12}. A well-known example of chaotic behavior in a first-order time-delay system is the Mackey-Glass oscillator describing the concentrations of chemicals in blood \cite{mackey77,zunino12,gao06}. According to the Takens theorem, see e.g.\ \cite{packard+80,wangj+16}, the length of the time-delay should be compatible with the embedding dimension. Consequently, one should expect sufficiently long time-delay as a signature of hyper-chaos to unveil chaotic behavior.

In this \textit{Letter} we account for the chaotic pulsar spin-down behavior by making use of nonlinear superfluid dynamics with taking glitch effects into consideration as a time-delay to the spin-down law.  As we show below, nonlinear superfluid coupling to the normal matter crust and in turn to the dipolar spin-down provides the required set of equations [c.f.\ Eqs. (\ref{nlcreep}) and (\ref{eq:lagdyn})], and glitches introduce a time-delay into spin evolution dynamics.

\textit{Nonlinear superfluidity and glitches.--}
Neutron stars can be considered as a two component system in which crust and superfluid couple to each other through friction maintained by the motion of the vortex lines. The resulting coupling is an important ingredient for the description of the superfluid dynamics as it determines the interaction time-scale between superfluid and normal (non-superfluid) matter and rotational irregularities in neutron stars like glitches, timing noise and precession \cite{haskell18}. In neutron stars, vortex lines exist in a very inhomogeneous medium and can pin to lattice nuclei in the crust  \cite{anderson75} and to magnetic flux tubes in the outer core \cite{erbil14}. In the vortex creep model \cite{alpar84,alpar89} vortex motion is described in terms of thermal flow under pinning forces and radial bias provided by the pulsar spin-down. As a result of spontaneous magnetization of the core superfluid's vortex lines, the neutron star core, which involves most of the moment of inertia of the neutron star, is effectively involved in the crustal normal matter, i.e.\ $I_{\rm c}\cong I$ \cite{als84}. Therefore, the only stellar component loosely coupled to the crust is the crustal superfluid. In the steady state, angular velocity lag $\omega\equiv\Omega_{\rm s}-\Omega_{\rm c}$  between the crustal superfluid and crust rotation rates is such that both components slow down by the same amount. 

Following G\"{u}gercino\u{g}lu and Alpar \cite{erbil17} we write the time evolution of the spin-down rate of the crust and  angular velocity lag dynamics as
\begin{align}
&\dot\Omega_{\rm c}= \frac{N_{\rm ext}(t)}{I_{\rm c}} + \sum_{i=1}\frac{I_{\mathrm{s},i}}{I_{\rm c}}\frac{\varpi_i}{2\tau_{\mathrm{l},i}}\exp(\omega/\varpi_i),
\label{nlcreep}\\
&\dot{\omega}=-\sum_{i=1}\frac{I \,\,\varpi_i}{2 I_{\rm c}  \tau_{\rm l,i}} \exp(\omega/\varpi_i)-\frac{N_{\rm ext}(t)}{I_{\rm c}},  
\label{eq:lagdyn}
\end{align}  
with $I_{\rm cs}=\sum I_{\mathrm{s},i}$ is crustal superfluid moment of inertia. The external torque $N_{\rm ext}(t)=-[\mu^{2}\Omega_{\rm c}(t)^{3}/c^{3}](1+\sin^{2}\alpha)$ depends on the magnetic dipole moment $\mu$, speed of light $c$, and inclination angle between the rotation and magnetic axes, $\alpha$ \cite{spitkovsky06}. Time dependence of $N_{\rm ext}(t)$ arises from $\Omega_{\rm c}(t)$. $\varpi\equiv (kT/E_{\rm p})\omega_{\rm cr}$, linear superfluid coupling time $\tau_{\rm l}$ and nonlinear superfluid recoupling time $\tau_{\rm nl}\equiv \varpi/|\dot\Omega|$ all depend on the pinning energy $E_{\rm p}$, crustal temperature $kT$ and critical lag for unpinning $\omega_{\rm cr}$ \cite{alpar89}. The ratio $\eta\equiv\tau_{\rm l}/\tau_{\rm nl}$ determines the type of the superfluid response. If $\eta > 1$, superfluid is in the nonlinear response regime. As can be seen from \autoref{tab:modelpar} most part of the crustal superfluid is in the nonlinear regime.  

For laminar superfluid flow, the typical distance between consecutive vortex lines $\ell_{\rm v}=(2\Omega_{\rm s}/\kappa)^{-1/2}\approx3\times10^{-3}(\Omega_{\rm s}/\mbox{\rads})^{-1/2}$ cm is much larger than the line's width, i.e.\ coherence length $\xi\approx10$ fm. Due to the presence of attractive pinning sites, i.e.\ lattice nuclei, vortex lines acquire curved shape \cite{chau93} as well as feel differential pinning potential \cite{cheng88} which will result in proximity effects for collective unpinning \cite{warszawski12}.  

A glitch occurs wherever and whenever equilibrium lag raises to the critical threshold. At the time of a glitch vortex lines in a particular crustal superfluid region unpin from lattice nuclei and impart their quantized angular momentum to the crust, thereby spinning up the crust. Number  of vortices discharged in a glitch, $N_{\rm v}$, is related to the change $\delta\Omega_{\rm s}$ in the crustal superfluid rotation rate by $N_{\rm v}=2\pi R^{2}\delta\Omega_{\rm s}/\kappa$ with $R$ and $\kappa$ being the neutron star radius and vorticity attached to each vortex, respectively. Glitches can be thought as sudden perturbations to the pulsar spin parameters such that rotation rate increaes by $\Delta\Omega_{\rm c}$ and the steady state lag decreases by $\delta\omega=\Delta\Omega_{\rm c}+\delta\Omega_{\rm s}$, i.e.\ $\Omega_{\rm c}(0)\rightarrow\Omega_{0}+\Delta\Omega_{\rm c}$ and $\omega(0)\rightarrow\omega_{\infty}-\delta\omega$. By angular momentum conservation $I_{\rm c}\Delta\Omega_{\rm c}=I_{\rm cs}\delta\Omega_{\rm s}$. Analyses reveal that the number of vortices involved in glitches is remarkably at the same order of magnitude and related to the broken plate size and in turn to the critical strain angle of the solid crust \cite{akbal18,erbil19}. This can be a clue for the future research on invariant properties of the pulsars.

If we assume that whole crustal superfluid behave as one single component following a glitch, then the immediate post-glitch spin-down and rotation rates becomes
\begin{align}
\dot\Omega_{\rm c}(t)=\frac{N_{\rm ext}(t)}{I_{\rm c}}+\frac{I_{\rm s}}{I_{\rm c}}\dot\Omega_{0}\left[1-\frac{1}{1+({\rm e}^{t_{\rm g}/\tau_{\rm nl}}-1){\rm e}^{-t/\tau_{\rm nl}}}\right],
\label{delayw}
\end{align}
and
\begin{align}
\Omega_{\rm c}(t)=\Omega_{0}(t)+\Delta\Omega_{\rm c}+\frac{N_{\rm ext}(t)}{I_{\rm c}}t+\frac{I_{\rm s}}{I_{\rm c}}\varpi \ln \{...\},
\end{align}
\label{delaywdot}
respectively, where
\begin{align}
\ln \{...\}=\ln \Bigg\{ 1+\frac{I}{I_{\rm c}}\exp\left(-\frac{t_{\rm g}}{\tau_{\rm nl}}\right)\left[\exp\left(\frac{t}{\tau_{\rm nl}}\right)-1\right]\Bigg\}.
\end{align}
We use the notation subscript ``0'' to describe the value of the corresponding quantity for the glitch-free case. After offset time $t_{\rm g}=[\Delta\Omega_{\rm c}+\delta\Omega_{\rm s}]/|\dot\Omega|$ elapses, pulsar resumes the steady spin properties towards the original pre-glitch state, $\Omega_{\rm c}(t)=\Omega_{0}+[N_{\rm ext}/I_{\rm c}]t$. So, the offset time $t_{\rm g}$ introduces a delay in the dynamical behavior of the pulsar through the factor $\exp(t-t_{\rm g})$. This delay-time cannot be known for each glitch, but it can be predicted.
\begin{figure}
    \begin{center}
        \leavevmode
        \includegraphics[width=0.9\columnwidth]{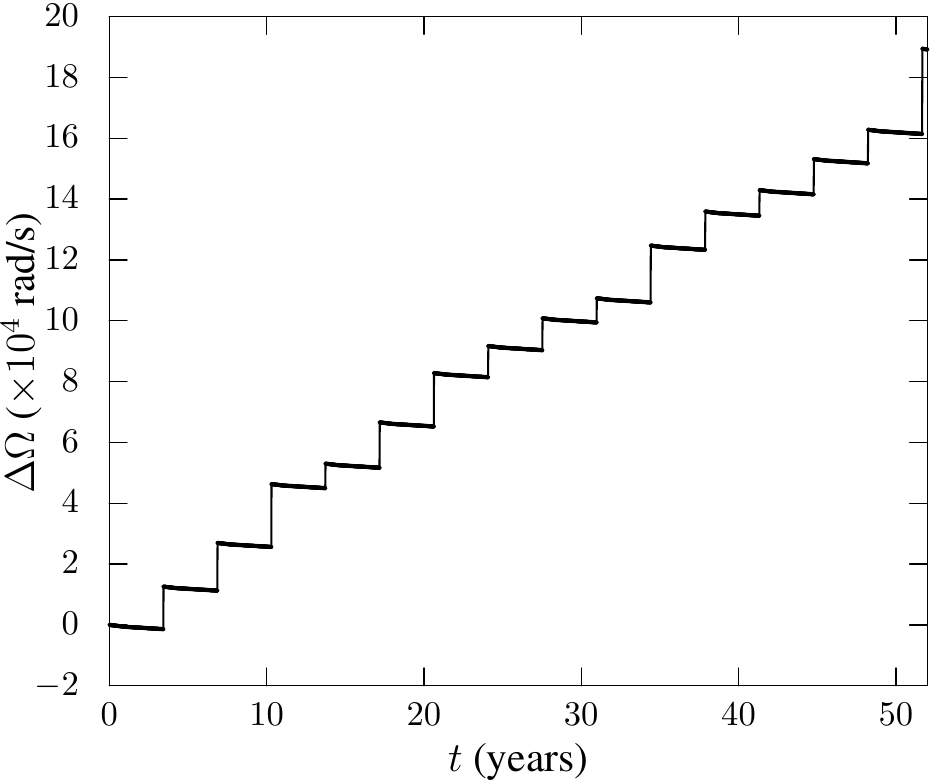}
    \end{center}
    \caption{
Residual of the rotation rate $\Delta\Omega$ after a series of glitches.
		}
    \label{residual1}
\end{figure}

\textit{Physics input.--}
The Vela pulsar is an emblematic source among glitching pulsars with large glitches of $\Delta\Omega_{\rm c}/\Omega_{\rm c}\approx2\times10^{-6}$ repeating in every 2-3 years \cite{manchester69,cordes88,akbal17,erbil21}. 
It has undergone 16 large glitches since 1969. 
We use physical parameters pertaining to this pulsar. 
We adopt $\alpha=85^\circ$ as determined from gamma-ray light curve \cite{rookyard15}. 
The vortex creep model parameters $\varpi$, $\tau_{\ell}$, and $\tau_{\rm nl}$ corresponding to the Vela pulsar for five crustal pinning layers \cite{erbil20} are given in \autoref{tab:modelpar}. We consider APR equation of state (EoS) \cite{akmal98} for computation of $I_{\rm s}/I_{\rm c}$ ratios corresponding to five pinning layers in \autoref{tab:modelpar}. We employ 1.424M$_{\odot}$ mass APR EoS which gives $I = 1.65\times10^{45}$\gcs~and $I_{\rm cs}/I_{\rm c}=3.65\times10^{-2}$ for total stellar and fractional crustal superfluid moments of inertia after correcting core superfluid's response \cite{erbil17a}, respectively. We run our code for a time-range of 52 years with current spin parameters of the Vela pulsar in order to make a direct comparison to the observational data. 

\begin{table}
\caption{Vortex creep model parameters for five pinning regions in the neutron star inner crust (for the Vela pulsar) \citep{erbil20}. Pinning energy calculations of Ref.\cite{seveso16} ($\beta=3$ model) are used  throughout.}
\label{pinpar}
\begin{center}{\tiny
\begin{tabular}{cccccc}
\hline\hline\\
\multicolumn{1}{c}{$\rho$} & \multicolumn{1}{c}{$\varpi$} & \multicolumn{1}{c}{$\tau_{\ell}$} & \multicolumn{1}{c}{$\tau_{\rm nl}$} & \multicolumn{1}{c}{$\omega_{\rm cr}$} & \multicolumn{1}{c}{$\omega_{\infty}$} \\
\multicolumn{1}{c}{($10^{13}\mbox{g cm$^{3}$}$)} & \multicolumn{1}{c}{($\mbox{rad s$^{-1}$}$)} & \multicolumn{1}{c}{(days)} & \multicolumn{1}{c}{(days)} & \multicolumn{1}{c}{($\mbox{rad s$^{-1}$}$)} & \multicolumn{1}{c}{($\mbox{rad s$^{-1}$}$)}\\ 
\hline\\
0.15 & 2.99$\times10^{-3}$ &2.29$\times10^{4}$ & 354 &9.67$\times10^{-2}$ &1.66$\times10^{-2}$ \\\\
0.96 & 3.31$\times10^{-4}$ & 3.56$\times10^{8}$ & 39 &1.48$\times10^{-2}$ &5.77$\times10^{-3}$ \\\\
3.4 & 2.76$\times10^{-4}$ & 7.27$\times10^{171}$ & 33 &1.16$\times10^{-1}$ &1.09$\times10^{-1}$ \\\\
7.8 & 9.97$\times10^{-5}$ & 1.87$\times10^{36}$ & 12 &1.10$\times10^{-2}$ &$8.22\times10^{-3}$ \\\\
13 & 1.38$\times10^{-4}$ &1.18$\times10^{13}$ & 16 &7.96$\times10^{-3}$ &3.97$\times10^{-3}$\\\\
\hline\\
\label{tab:modelpar}
\end{tabular}}
\end{center}
\end{table}

\textit{Results.--}
\begin{figure}
    \begin{center}
        \leavevmode
        \includegraphics[width=0.9\columnwidth]{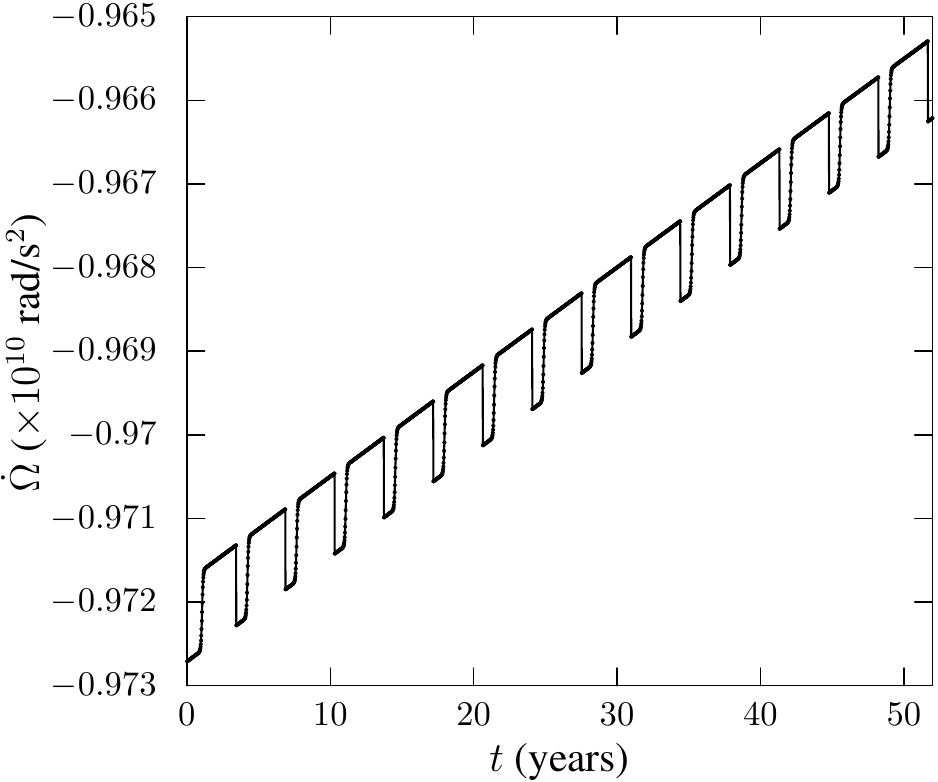}
    \end{center}
    \caption{The behavior of the spin-down rate $\dot\Omega$ after a series of glitches.
    }
    \label{residual2}
\end{figure}
We numerically solve the coupled Eqs. (\ref{nlcreep}) and (\ref{eq:lagdyn}) with reset condition that a glitch occurs whenever spin-down rate drives the angular velocity lag to the critical threshold
 \begin{align}
 |\dot\Omega_{\rm c}(t)|\times T=\omega_{\rm cr},
 \label{gcond}
 \end{align}
where $T$ is the time since the previous glitch. We assume that the vortex unpinning avalanche starts at the densest pinning region in the crust, i.e.\ from the fifth layer in \autoref{tab:modelpar}, and thereafter covers the whole crustal superfluid. At each glitch we introduce perturbations to both rotation and spin-down rates. We choose the observed distribution for the Vela pulsar, $\langle\Delta\Omega_{\rm c}/\Omega_{\rm c}\rangle=2.04\times10^{-6}$ with deviation $\sigma_{\Delta\Omega_{\rm c}/\Omega_{\rm c}}=0.77\times10^{-6}$. For the lag value at the time of glitch $\omega(0)\cong\omega_{\infty}-(1+I_{\rm c}/I_{\rm cs})\Delta\Omega$, we consider distribution corresponding to the observed glitch magnitudes of the Vela pulsar with mean $3.20\times10^{-2}\omega_{\rm cr}$ and deviation $1.44\times10^{-2}\omega_{\rm cr}$, where $\omega_{\rm cr}=7.96\times10^{-3}$\rads~is adopted from the densest pinning region [c.f.\ \autoref{tab:modelpar}].
 
 In \autoref{residual1} we show residual $\Delta\Omega=\Omega_{\rm glitch}-\Omega_{\rm glitch~free}$ of the rotation rate after a series of glitches for our numerical setup. Staircase increments are evident. Notice that our synthetic results are very similar to the observed record of the Vela pulsar \cite{andersson12}.
 
  In \autoref{residual2} we show the spin-down rate after a series of glitches for our numerical setup. Note that our synthetic results are very similar to the observed behavior of the Vela pulsar wherein glitch induced spin-down increase recovers almost completely before the next glitch arrives. 

\begin{figure}
    \begin{center}
        \leavevmode
        \includegraphics[width=0.9\columnwidth]{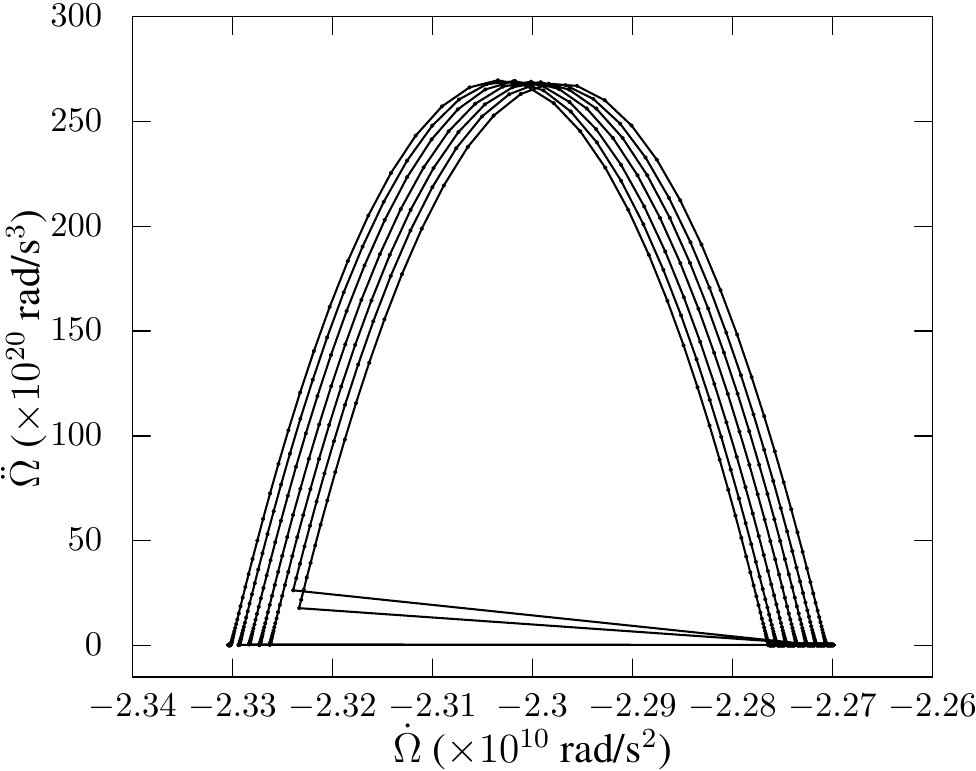}
    \end{center}
    \caption{Glitch induced residual in the second time derivative of the rotation rate.}
    \label{wddotcor}
\end{figure}

In \autoref{wddotcor} we show the correlation between $\ddot\Omega_{\rm c}$ and $\dot\Omega_{\rm c}$. For visualisation purposes we plot for the case with 7 glitches occurred in 10 years in which $\mu$ is increased by a factor of 1.55. The pattern is reminiscent of the M\"{o}bius strip. Biryukov et al.~\cite{biryukov12} attributed the cyclic behavior of $\ddot\Omega_{\rm c}$ to a cyclic spin-down component underlying the magneto-dipole radiation. But as can be seen from our results it can be a natural outcome of unresolved glitches in the long-term timing data of older pulsars.
Moreover, time-delay introduced by unresolved glitches may cause dual spin-down characteristics observed by Ref.~\cite{lyne10}.

In the glitch-free case the rotational phase of a pulsar can be evaluated by a Taylor series expansion up to the third derivative and recast in the form
\begin{align}
\Phi(t)_{\rm glitch~free}=\Phi_{0}+\Omega_{0}t+\frac{\dot\Omega_{0}}{2}t^2+\frac{\ddot\Omega_{0}}{6}t^3.
\end{align}

\begin{figure}
    \begin{center}
        \leavevmode
        \includegraphics[width=0.9\linewidth]{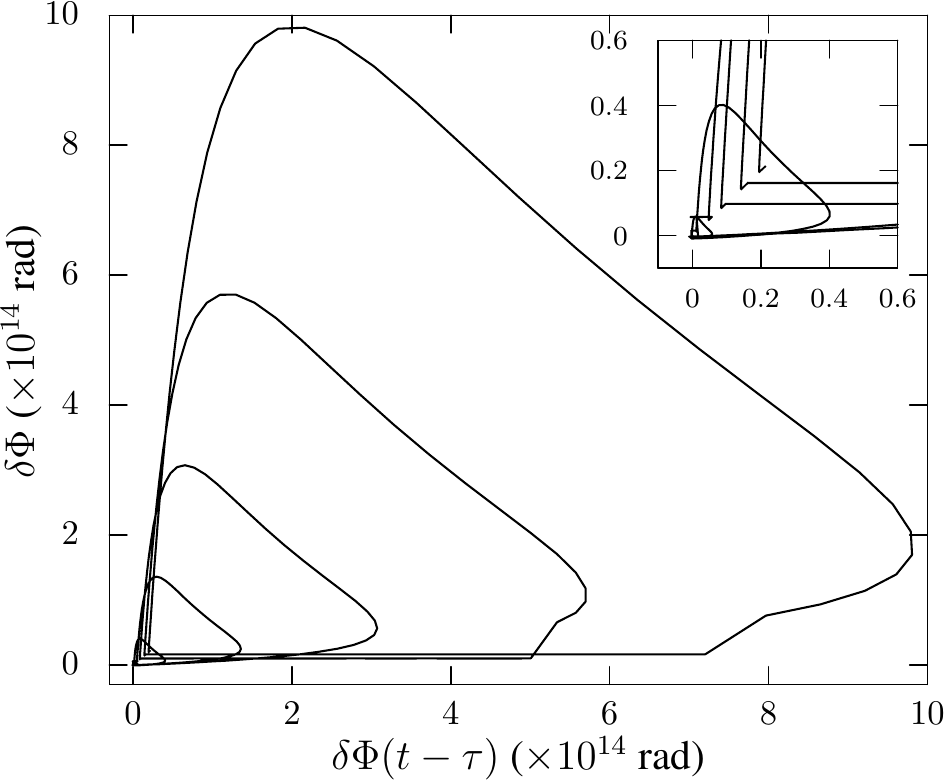}
    \end{center}
    \caption{Phase difference between glitch and glitch-free cases for a time-delay of $\tau=60$ days. Inset shows the sixth cyle.
   		}
    \label{phasedif}
\end{figure}

\begin{figure}
    \begin{center}
        \leavevmode
        \includegraphics[width=0.9\linewidth]{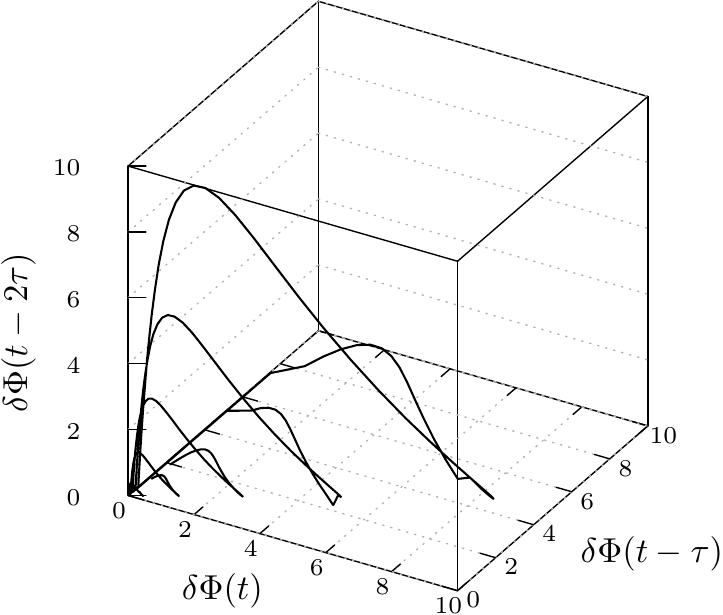}
    \end{center}
    \caption{Three dimensional delayed phase difference between glitch and glitch-free cases corresponding to $\tau=60$ days.
   		}
    \label{phasedif3d}
\end{figure}

We composed phase difference $\delta\Phi=\Phi_{\rm glitch}-\Phi_{\rm glitch~free}$ by using our numerical setup for glitches. We find that chaotic structures emerge in the time-delayed phase difference diagram starting from a delay-time of $\tau=40$ days and become optimum at $\tau=80$ days in accordance with the fact that the existence of a time-delay with significant duration in a system renders it infinite-dimensional. Notice that this delay time-scale $\tau$ is of order of nonlinear creep recoupling time $\tau_{\rm nl}$, see \autoref{tab:modelpar}. In \autoref{phasedif} we plot $\delta\Phi(t)$ against $\delta\Phi(t-\tau)$ for a delay-time of $\tau=60$ days again for the case with 7 glitches occurred in 10 years. Despite we did not use real observed data, our synthetic data for $\delta\Phi(t)=\Phi(t)_{\rm glitch}-\Phi(t)_{\rm glitch~free}$ captures the essential characteristic of the corresponding figure 1 of Harding et al. \cite{harding90} who used real timing data. In \autoref{phasedif3d} we show three dimensional time-delayed phase difference diagram in which higher dimensional, cusp-like structures are evidence for hyper-chaos comparable with the ones in \cite{junges+12,zunino12,packard+80}. We note that our findings do not rely on the presence of 5 pinning regions since we were able to obtain similar results involving chaos with a single region by small changes in the model parameters.


\textit{Discussion.--} 
In this \textit{Letter} we have investigated nonlinear superfluidity and glitch effects on the spin evolution of pulsars. 
We solved for the coupled set of equations of a two-component neutron star system in which the coupling is mediated by the motion of superfluid vortex lines under the combined action of pinning potential of various strength and spin-down. 
We elaborated on the glitch perturbations to the steady-state dynamics and observed that glitches introduce a delay of timescale $t_{\rm g}$ to the dynamical state of the crustal rotation, 
an important ingredient for the chaotic behavior seen from some pulsars. 
We emphasize that the time-delay based hyper-chaos we conjectured in this \textit{Letter} is fundamentally different than the classical approach for chaotic attractors wherein three-dimensional continuous systems are considered \cite{seymour13,harding90}.

With this simplistic model we reproduced the glitching behavior of the prolific Vela pulsar. By theoretically examining the lag between the superfluid and crustal rotation rates during a glitch, and using reasonable values for the pulsar parameters, we obtained that spin residuals display staircase increments (see \autoref{residual1}) while the spin-down rate following a glitch decays with a constant, larger $\ddot\Omega_{\rm c}$ and recovers almost completely before the next glitch arrives (see \autoref{residual2}). We were also able to obtain 16 large glitches in a timespan of 52 years. All of these features characterise well the observed glitch properties of the Vela pulsar \cite{alpar84,erbil21,cordes88,akbal17}.

Decoupling of the vortices during a spin-up glitch results in acceleration of the pulsar with $\ddot\Omega_{\rm c}=|\Delta\dot\Omega_{\rm c}|/t_{\rm g}$, which is about two to four orders of magnitude higher than the steady-state value. Naturally, glitch-induced transient changes in $\ddot\Omega_{\rm c}$ are in the form of poles and leave behind a residual in the rotational-phase data. Hence, we can use alternation theorem \cite{vaidyan07} to invent a relation between the number of glitches and the residual phase, see \autoref{phasedif}. Our study here introduces a new method to determine the actual number of glitches suffered by a pulsar even in the cases of low observational cadence and huge data gaps. If our conjecture is true, one can quantify the actual number of glitches of a pulsar by producing the time-delayed rotational phase-difference diagram and then counting the number of cycles in the diagram. The number of cycles is equal to one-less of the total number of glitches underwent by the pulsar for the corresponding observation time. We have studied this relation for different number of glitches and seen that it holds for in all cases. \autoref{phasedif} shows an example.

The observed chaotic behavior for pulsar spin-down is similar to the hyper-chaos induced by Mackey-Glass oscillator; see the cusps in \autoref{phasedif3d} and e.g.\  figures 2 and 3 in Ref. \cite{junges+12}. Analogous to the Mackey-Glass system which deals with the delayed response to the concentration variation of a substance in blood flow around cells, pulsar spin-down after a series of glitches proceeds as delayed-response to the changes in the rates of internal superfluid friction and angular momentum transfer to the crust due to the disturbance in the flow or motion of the vortex lines.    

\begin{acknowledgments}
EG acknowledges support from the Scientific and Technological Research Council of Turkey (T{\"U}B{\. I}TAK) with grant number 117F330.
KYE acknowledges support from T{\"U}B{\. I}TAK with grant number 118F028. We thank M.~Ali Alpar for useful comments on the paper. 
\end{acknowledgments}

\bibliography{literature}

\end{document}